\documentclass{ws-ijmpb}

\begin{document}

\markboth{G. P. Tsironis}
{Nonlinear dimer}

%%%%%%%%%%%%%%%%%%%%% Publisher's Area please ignore %%%%%%%%%%%%%%%
%
\catchline{}{}{}{}{}
%
%%%%%%%%%%%%%%%%%%%%%%%%%%%%%%%%%%%%%%%%%%%%%%%%%%%%%%%%%%%%%%%%%%%%

\title{Dynamical symmetry breaking through AI: The dimer self-trapping transition \\  }

\author{G. P. Tsironis}

\address{Institute of Theoretical and Computational Physics and Department of Physics, University of Crete,  P.O. Box 2208, 71003 Heraklion, Greece\\
gts@physics.uoc.gr}

\author{G. D. Barmparis}

\address{Institute of Theoretical and Computational Physics and Department of Physics, University of Crete,  P.O. Box 2208, 71003 Heraklion, Greece\\ 
barmparis@physics.uoc.gr}

\author{D. K. Campbell}

\address{Physics Department, Boston University, MA 02215, USA\\
dkcampbe@bu.edu}

\maketitle

\begin{history}
\received{\today} % Day Month Year}
%\revised{Day Month Year}
%\accepted{(Day Month Year)}
%\comby{(xxxxxxxxxx)}
\end{history}

\begin{abstract}
The nonlinear dimer obtained through the nonlinear Schr{\"o}dinger equation has been a workhorse for the discovery the role nonlinearity plays in strongly interacting systems.  While the analysis of the stationary states demonstrates the onset of a symmetry broken state for some degree of nonlinearity, the full dynamics maps the system into an effective $\phi^4$ model.  In this later context the self-trapping transition is an initial condition dependent transfer of a classical particle over a barrier set by the nonlinear term.  This transition has been investigated analytically and mathematically it is expressed though the hyperbolic limit of Jacobian elliptic functions. The aim of the present work is to recapture this transition through the use of methods of Artificial Intelligence (AI). Specifically, we used a physics motivated machine learning model that is shown to be able to capture the original dynamic self-trapping transition and its dependence on initial conditions.  Exploitation of this result in the case of the non-degenerate nonlinear dimer gives additional  information on the more general dynamics and helps delineate linear from nonlinear localization. This work shows how AI methods may be embedded in physics and provide useful tools for discovery.

\end{abstract}

\keywords{Nonlinear Schr{\"o}dinger Equation; self-trapping transition; elliptic functions; nonlinear dimer; physics motivated machine learning.}

\section{Introduction}
The life of the famous Discrete Nonlinear Schr{\"o}dinger (DNLS) equation started with a different name, i.e. as the Discrete self-trapping (DST) equation introduced by Eilbeck, Lomdhal and Scott in 1985 in a seminal paper that marked nonlinear dynamics\cite{eilbeck}.  The DST equation was motivated by biology and the idea of the Davydov soliton that was thought to dominate the energy transfer processes in proteins\cite{scott}.  In this first paper Eilbeck et al. showed that in the stationary version of the DST generates an abundance of bifurcations and states produced through nonlinearity that grow in numbers and complexity as the number of units increases.  The simplest case of two units, i.e. the dimer, although it also shows the emergence of nonlinear states it is fully integrable while systems with larger number of units are generally non-integrable and chaotic.  The complete dynamical analysis of the DNLS dimer was performed soon after this by Kenkre, Campbell and Tsironis where the onset of the self-trapping transition is seen as strongly initial-condition dependent passage over a barrier\cite{kenkre,ktc,gpt1}.  The presence of nonlinearity turns the linear dimer trigonometric evolution to elliptic function evolution while the self-trapping transition itself is nothing but the reduction of elliptic functions to hyperbolic ones.  Physically, nonlinearity slows down the transfer from one site of the dimer to the other in a process that takes infinite amount of time at a critical nonlinearity while  incomplete transfer marks the self-trapping regime\cite{kenkre}.   These results are important for strongly interacting electron-phonon systems of molecular crystals but the DNLS equation appears also in photonics.  In this context Christodoulides and Jospeh analyzed the two optical fiber nonlinear dimer and showed that self-trapping may assist in the design of fiber systems with designed switching properties\cite{christodoulides}.  The work mentioned already on the dimer focused on the degenerate case where both units are identical; if they are not we may have an energy mismatch build in.  In this non-degenerate nonlinear dimer case the role of the self-trapping is mixed with the energy mismatch and although the system remains integrable and solvable through elliptic functions the behavior is more complex\cite{gpt2}.

The DNLS equation nonlinear dimer is a remarkable system that is simple enough to be studied analytically yet it contains non-trivial complexity.  For the sake of pictorial simplicity let us assume we have two molecular units that each have one available energy state\footnote{We use the "condensed matter picture" here. Similar interpretation follows in the "photonics" picture that we presently avoid for simplicity.}  When both states have the same energy, the dimer is degenerate and the wavefunction overlap $V$ determines the transfer time from one site to the second.  In this case the transfer is complete in the sense that an initial excitation placed fully on site one may transfer completely to the second site.  The transfer occurs because the two sites are at resonance-since they have the same energy value- and, in this case, the transfer element $V$ facilitates this resonant transfer.  When the energies are not the same, as in the case of the non-degenerate dimer, the two sites cannot be fully resonant.  In this system  the matrix element $V$ only transfers part of the excitation to the second site; clearly the amount transferred depends on the competition between the energy mismatch $\Delta$ and the transfer $V$.  This linear picture transfers to some extent in the nonlinear dimer case.  When we have the fully degenerate dimer, the nonlinearity affects equally both sites leading to reduction of excitation transfer speed yet still enabling full transfer at small nonlinearity values.  At a given critical nonlinearity symmetry breaking occurs and complete transfer is replaced by incomplete transfer to the second site.  In other words ,nonlinearity introduces dynamically an effective energy mismatch and renders the nonlinear dimer non-degenerate.  This trend increases with nonlinearity and for very large nonlinearities the transfer becomes very small\footnote{This picture is strongly initial condition dependent and it can be altered if initial phases are introduced \cite{gpt1}}.  

If strong nonlinearity turns the degenerate dimer into an effectively non-degenerate one then when we start with a non-degenerate dimer we expect the two tendencies to augment. In fact for localized initial conditions the non-degeneracy introduces resonance mismatch while nonlinearity generally augments this tendency and the transfer is even less complete.  There is however an exceptional case where nonlinearity acts in such as away as to eliminate effectively the energy mismatch introduced by nondegeneracy.  This is the case of Targeted Energy Transfer (TET) introduced by Kopidakis, Aubry and Tsironis, where it was shown analytically that appropriate choice of nonlinearity restores the ultra-fast transfer of the purely linear degenerate dimer\cite{kopidakis}.  This somehow surprising behavior in TET comes however with a price, viz. the nonlinearity in the two dimer sites has the same absolute value but opposite sign.  One site is thus attractive and the other repulsive; the last feature could stem physically from a capacitive effect induced from local charge accumulation.  The discovery of TET completes in some conceptual sense  that path that starts from the linear degenerate dimer: The resonant transfer that is inhibited by either linear non-degeneracy or nonlinear self-trapping is fully restored in the TET dimer that encompassed both in an appropriate way.

The aim of the present work is to investigate if some of the discoveries in the nonlinear dimer outlined previously may be addressed through AI tools.  More specifically, we would like to know if Machine Learning (ML) motivated from physics may play some role in discovering symmetry breaking properties of the nonlinear dimer.  We believe that this is an interesting question since the dimer is well studied analytically and thus it can form some sort of test bed for these methods.  If successful then can be applied to other more complex cases.  The more specific target of this work is the self-trapping transition, the landmark of the nonlinear dimer and whether it can be predicted by ML. 

The structure of the paper is the following: In the next section, we introduce the math of the DNLS equation, describe explicitly the self-trapping transition and give quantitative information.  In section 3, we use ML and describe the predictions related to self-trapping. Here we detail our ML method and give the results for the degenerate nonlinear dimer both for localized and more general initial conditions. In section 4, we focus on the non-degenerate nonlinear dimer, describe its dynamical phase diagram and show how ML can capture the transition here as well.  Finally, in section 5, we conclude and provide a more general AI-based picture on the dimer studies.

\section{The nonlinear dimer}

In this section we review basic nonlinear dimer properties.  We start from a general expression of the DNLS equation written in the form 

\begin{equation}\label{EQ-1}
i \frac{d \psi_n}{dt} = \epsilon_n \psi_n + V( \psi_{n+1} + \psi_{n-1} ) - \chi_n |\psi_n |^2 \psi_n
\end{equation}

where, $\psi_n \equiv \psi_n (t)$ is a complex variable at time $t$ while $\epsilon_n $, $\chi_n$ and $V$ are parameters of the problem.  We will follow the condensed matter interpretation of the equation\cite{hennig}. In this representation we can think of a one-dimensional infinite lattice where each site is labeled by the index $n$, we have local site energy $\epsilon_n$ and local nonlinearity $\chi_n$ while $V$ is the common nearest-neighbor integral overlap.  A quantum mechanical particle tunnels from site to site while experiencing a nonlinear interaction due to strong coupling with other degrees of freedom-in the LHS of the equation we have suppressed $\hbar$.  The complex quantity $\psi_n $ is simply the probability amplitude for the particle to be found at the $n$-th unit.

For the dimer we have only two units; we note the occupation probability difference between the two sites with $p(t) = | \psi_1 (t) |^2 - | \psi_2 (t) |^2 $ as in Ref. \cite{kenkre}. We designate further the energy difference between the energy levels as $\Delta = \epsilon_2 - \epsilon_1$; with the exception the case of TET both nonlinear parameters are equal to each other, i.e.  $\chi_1 = \chi_2 = \chi$. It is useful to scale parameters as $\delta = \Delta / 2V$ and $\zeta = \chi / 4V$.  In the pure degenerate case ($\delta = 0$)  and for perfectly localized initial condition, i.e. $\psi_n (0) = \delta_{n,1}$ the self-trapping transition occurs at the critical value $\zeta_{cr} = 1$.  For $\zeta  < \zeta_{cr}$ there is complete transfer of the excitation between the two units although the motion becomes slower.  Specifically, the period of oscillation for $p(t)$ grows as $T=T_0 K( \chi / 4 V )$, where $K$ is the complete elliptic integral of first kind and $T_0 =\pi / 2V$ is the period of the linear dimer\cite{kenkre}.  For $\zeta > \zeta_{cr}$ the transfer becomes incomplete.  The reduction of transfer speed due to nonlinearity has been considered as a signature of polaronic effects while the incomplete transfer an effective introduction of an energy mismatch by nonlinearity.

In the linear non-degenerate dimer there is incomplete transfer from the beginning; here the nonlinearity accentuates this tendency.  It is noteworthy that at small non-degeneracies the increase of nonlinearity leads to a similar behavior that is characterized by a sudden decrease in the transfer to the other side, i.e. a behavior that may be similarly associated with a self-trapping transition\cite{gpt2}.  Non-degeneracy and nonlinearity work in a coordinated way but the nonlinearity has to reach a threshold value before we observe a qualitative change in the dynamics of transfer.  As mentioned earlier only in the TET case this cooperation of $\delta$ and $\zeta$ can be broken and the original resonant transfer restored.

\subsection{Methodology} \label{methodology}
The methodology in this work follows a path similar to the one proposed by Barmparis and Tsironis to discover nonlinear resonances through ML\cite{barmparis}. There is currently substantial  interest in ML approaches that utilize directly equations of physics or mathematics\cite{karniadakis}.   In the present work we integrate numerically  Eq. (\ref{EQ-1}) using a $4^{th}$ order Runge-Kutta method with an integration step of 0.005 and introduce a new data-free physics-informed loss function, designed to capture the desired properties of self-trapping transition defined as:

\begin{equation}
Loss(T^{min(P_1)}, \zeta) = \Big| 0.5 - P_1(T^{min(p_1)}, \zeta) \Big |
\label{loss}
\end{equation}
where, $P_1(T^{min(p_1)}, \zeta) = | \psi_{1}(T^{min(p_1)}, \zeta) |^2$ is the probability of the system being at state, $1$,  at time,  $T^{min(P_1)}$, for the given $\zeta$ value. This definition ensures that minimizing the loss function will conclude to a minimum occupation probability equal to 0.5 for state $1$.  It is equivalent to $P_1(T^{min(p_1)}, \zeta) - P_2(T^{min(p_1)}, \zeta)$, where $P_2 (t)$ is the occupation probability of the second site at the designated time $t$.  A difficulty one faces will training a ML model with the above loss function is that the minimum occupation probability at state $1$, $P_1$, becomes zero for all the values of $0 < \zeta < 1$. Thus the loss function ends up being flat and equal to 0.5. Having a flat loss function, independent of $ \zeta$, prohibits the model from updating its weights and thus the training process stops. This behavior is demonstrated in the area with the white background in Fig. \ref{self-trappingTraj}, where the trajectories initialized inside the range of $0 < \zeta < 1$ are stuck around the initial value of $\zeta$. In order to avoid having trajectories that result to parameters that do not satisfy the required properties, we initialize all of the trajectories to a large enough value of $ \zeta $.  Using a large initial value of $\zeta$ addresses the problem of having untrained trajectories, but it does not ensure the finding of the proper parameters, i.e. minimizing the loss function. The reason is that by using a large initial value for $\zeta$, we need to use a large value for the learning rate during training in order to accelerate the learning process. Using a large learning rate may result in substantial changes in the loss function and consequently significant changes to the value of $\zeta$, which may land to values inside the flat area of the loss function.  Thus, we introduce one more criterion to keep a large value for the learning rate and continue minimizing the loss function properly. This criterion checks if the minimum occupation probability, $P_1$, at the last step during the training process is less than 0.5. If this happens, it returns the value of $\zeta$ back to its previous value while reducing the learning rate by a factor of 10. This additional criterion ensures that we always approach the critical value of $\zeta$ from values  where $P_1(\zeta) > 0.5$ and that by using an adaptive learning rate, we will converge to the critical value. The algorithm was implemented in TensorFlow 2.4\cite{TF}/Keras\cite{Keras} using an Adam optimizer\cite{Adam} with a custom learning rate. The details of the algorithm can be found in Ref.\cite{barmparis}.

\section{Physics and Machine Learning in the nonlinear dimer}
In the nonlinear dimer  we have a complex dynamical problem that mixes three separate behaviors: The first is the  dynamical breaking of symmetry induced in the degenerate dimer case; this is the landmark of the self-trapping transition.  Nonlinearity becomes a dynamic mismatch agent and lifts the resonance between the two sites. The second involves nondegeneracy already at the linear regime that is accompanied by incomplete transfer.  In this case nonlinearity enhances the mismatch especially at large values.  Finally we can also restore resonance even in the non-degenerate case through the proper choice of nonlinearity; this is the case in TET. The question we pose now is whether we can detect this complex behavior through specific use of AI.  If the answer is affirmative this will lead to the development of a mechanism for the discovery of resonant properties of nonlinear dynamical systems.  Using the methodology outlined earlier we obtain the following results.
\\

\subsection{Localized initial conditions}
\begin{figure}[bt]
\centerline{\psfig{file=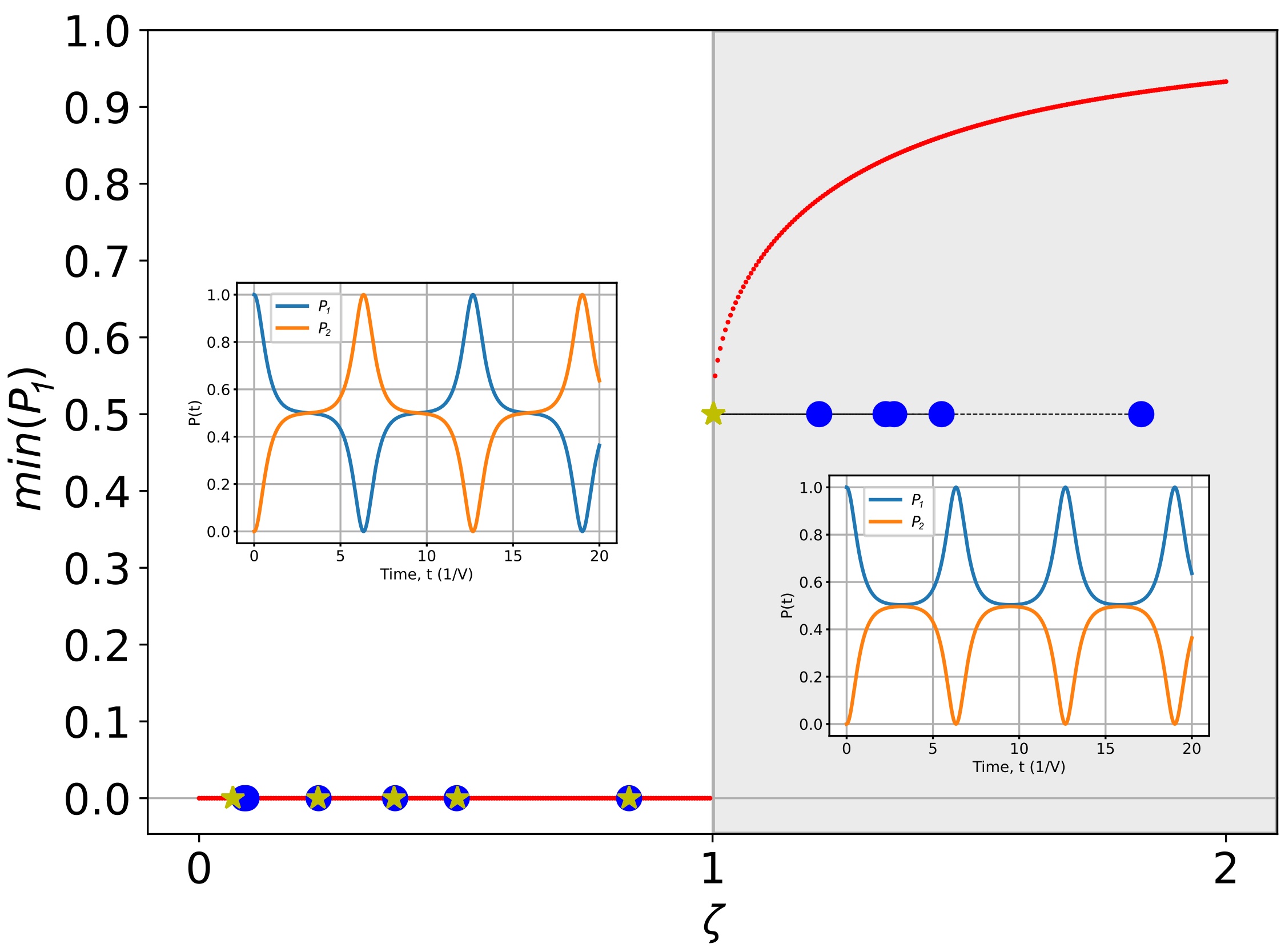,  width=5.65in}}
\vspace*{8pt}
\caption{The nonlinear dimer self-trapping transition with ML. The transition is obtained by following the minimal value of the occupation probability at the first site as a function of the scaled nonlinearity $\zeta$. The red dotted line denotes the calculated through ML actual occupation probability $P_1 (t)$.  This curve shows precisely the features of the self-traping transition.  The x-axis of the figure acts shows also the flow of trajectories for various initial conditions. The blue bullets denote the initial value of the nonlinearity parameter $\zeta$ of ten randomly initialized trajectories (black dashed lines), while the yellow star points to the final value of $\zeta$.  This final value is either the value that satisfies the desired properties of the system, i.e., self-trapping transition, or a position that the learning process was unsuccessfully stuck (See section \ref{methodology}). The area with the gray background ($\zeta > 1$) shows the range of values for the nonlinearity that conclude to self-trapping transition. In the inserts we show the time evolution across the transition for $\zeta \rightarrow 1_{-}$ (free motion) and $\zeta \rightarrow 1_{+}$ (self-trapped motion) respectively. }
\label{self-trappingTraj}
\end{figure}

Let us first give the results regarding the self-trapping transition starting from a localized initial condition.  This is shown in Fig. \ref{self-trappingTraj}, where the circles denote the initial value of the nonlinearity parameter $\zeta$ of several randomly initialized trajectories (black dashed lines), while the yellow star points to the final value of $\zeta$.  This value is either the value that satisfies the desired properties of the system, i.e.  self-trapping transition, or a position that the learning process was stuck as explained in sub-section \ref{methodology}. The red dotted line presents the minimum occupation probability at state 1,  $\min(P_1)$ as a function of the nonlinearity parameter $\zeta $.  The area with the gray background ($\zeta > 1$) shows the range of values for the nonlinearity that conclude to self-trapping transition.  The inner plots present the occupation probability of each state of the system as a function of time.  The plot inside the white background shows the situation just before the condition for self-trapping transition ($\zeta \rightarrow 1_{-}$) and the one inside the gray background the situation just after the self-trapping ($\zeta \rightarrow 1_{+}$).  We observe that the ML procedure we use discovers the self-trapping transition at $\zeta_{cr} = 1$, i.e. $\chi_{cr} = 4V$.

\subsection{General initial conditions}
\begin{figure}[bt]
\centerline{\psfig{file=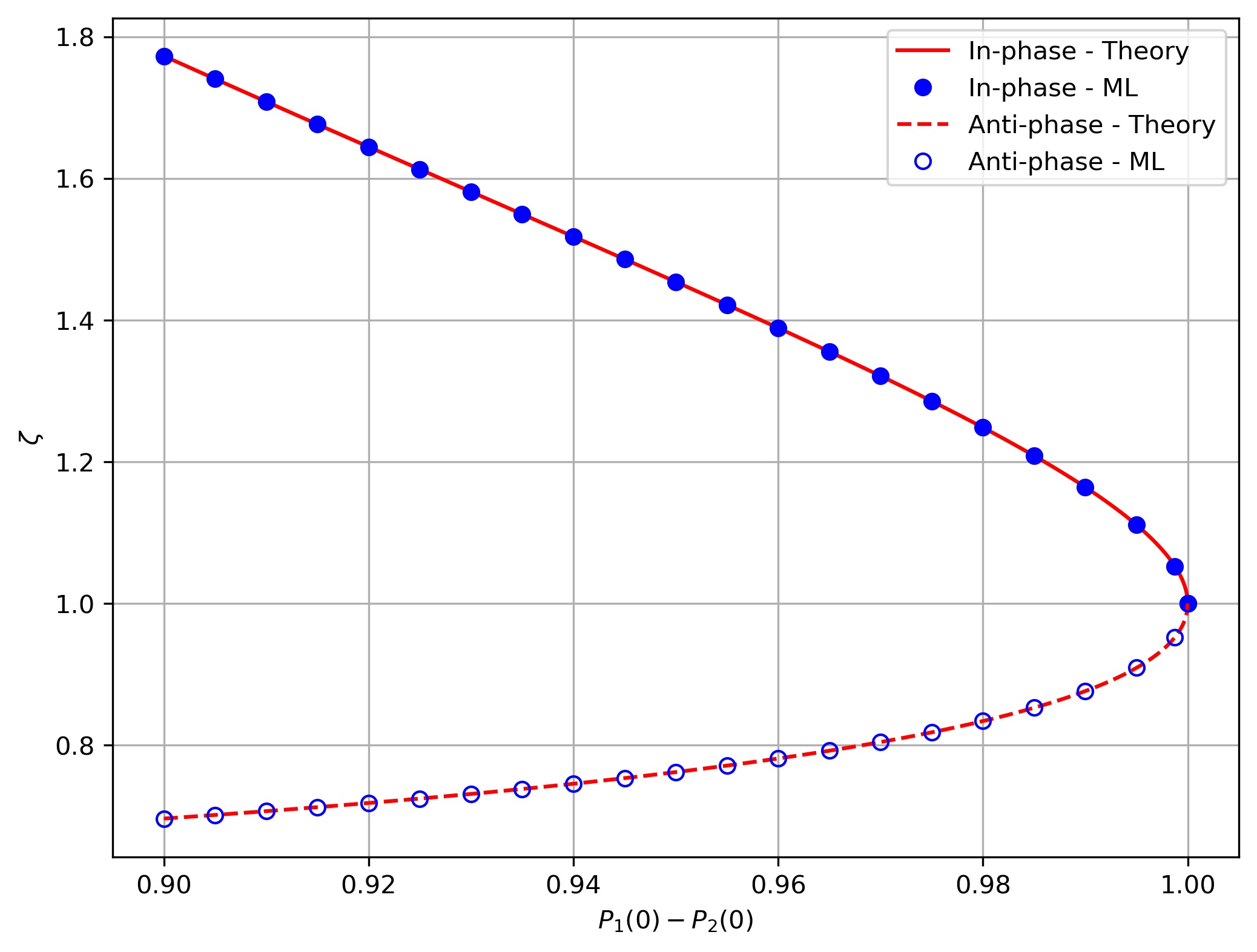,width=5.65in}}
\vspace*{8pt}
\caption{Self-trapping transition with ML for general initial conditions. The solid line is the curve of Eq. (\ref{EQ-2}) while the blue bullets were produced with ML. The two branches correspond to in-phase (solid red line) and anti-phase (dashed red line) initial conditions.}
\label{self-trappingVsInitialConds}
\end{figure}

When the initial placement of the excitation is not fully on one of the sites the amount of nonlinearity necessary for self-trapping changes.  For real off-diagonal elements of the density matrix $\rho_{mn} = \psi_m \psi_n^*$ the critical nonlinearity for self-trapping is
\begin{equation}\label{EQ-2}
\zeta_{cr} = \frac{1 \pm (1- p_0^2 )^{1/2}}{p_0^2}
\end{equation}
with $p_0 = \rho_{11} (0) - \rho_{22} (0)$ and where initial real amplitudes  may be ``in-phase" having the same sign or in ``anti-phase" with opposite sign\cite{gpt1}.  The former choice leads to the plus sign in Eq. (\ref{EQ-2}) while the latter to minus. We note that the critical nonlinearity for self-trapping increases as the amount of initial localization decreases for in-phase motion while the opposite for the out of phase motion.  In Fig. \ref{self-trappingVsInitialConds}, we show the results for general initial conditions with real and positive off-diagonal matrix elements of the density matrix. The continuous line is the analytical result of Eq. (\ref{EQ-2}) while the blue bullets present the results of the ML search.  Both in-phase and anti-phase branches are shown. We see a remarkable agreement between the two demonstrating that ML can discover faithfully arbitrary initial condition results in the dynamic phase transition.

\section{The Non-degenerate dimer}
In the non-degenerate nonlinear dimer (NNLD) nonlinearity mixes with the non-resonance condition induced by non-degeneracy.  We have two different aspects; one is the effective linear ``disorder" introduced by the energy mismatch.  The larger the energy difference the smaller the transfer of energy from one site to the next. The second feature is the nonlinearity; this introduces an effective mismatch as well that is however dynamic and controlled by the initial conditions.  Both non-resonant mechanisms act in the same direction, although the linear non-degeneracy is always present while nonlinearity operates better after a threshold value.  

The NNLD can be mapped into an effective problem of a particle in a potential well similar to the nonlinear dimer\cite{gpt2}.  Assuming localized initial conditions, in the latter the self-trapping transition occurs when the potential develops a flat region indicating the presence of a barrier separating the sites\cite{kenkre}.  If a similar criterion is applied in NNLD we find an interesting dynamic phase diagram where the transition point of the nonlinear dimer becomes a ``critical line"\cite{gpt2}. Furthermore, in addition to a linear-disorder dominated region as well as a nonlinear dominated one we also find a mixed region where the two tendencies mix.  The phase diagram is obtained through analytical means while the ML-based analysis that we now detail is completely independent of the mathematical approach. 

\begin{figure}[bt]
\centerline{\psfig{file=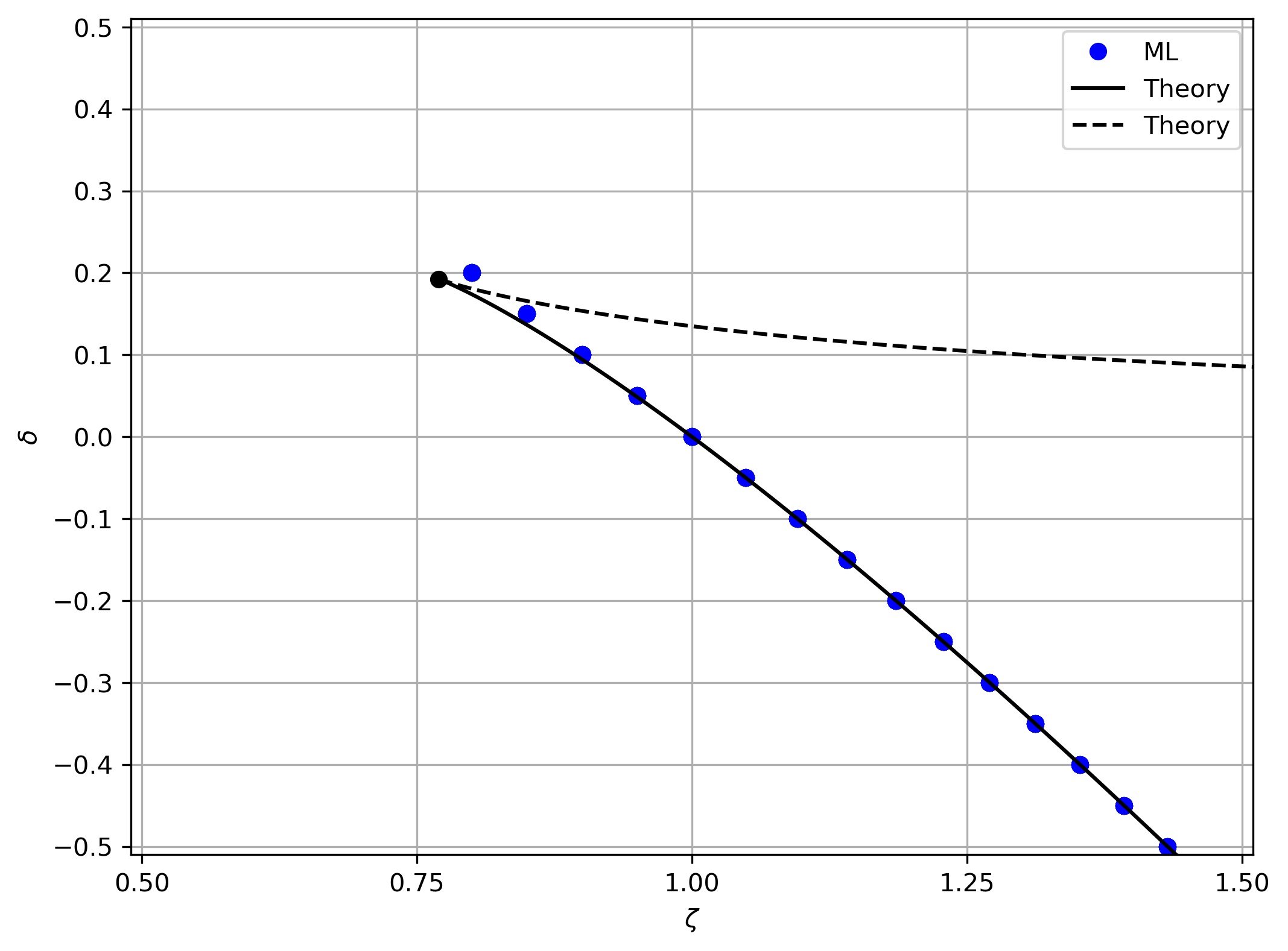,width=5.65in}}
\vspace*{8pt}
\caption{Self-trapping transition with ML for the non-degenerate nonlinear dimer with localized initial condition. The continuous lines are result of analytical calculations while the bullets a result obtained through ML. The solid continuous line represents the self-trapping transition that is a transition to a region of nonlinearly localized states. The ``tie" region between the continuous and dashed line corresponds to self-trapping or nonlinear localization.}
\label{NNLD}
\end{figure}

In the NNLD  for small non-degeneracies there is a sudden transition from quasi-resonant motion to selftrapped one\cite{gpt2}; this produces the solid line in Fig. \ref{NNLD}.  This line that replaces the self-trapping point of the degenerate case is fully captured by ML!  The second, dashed line, marks the end of the region where nonlinear localization dominates.  In the work of Archilla, MacKay and Marin  the distinction of linear versus nonlinear localization was discussed in the context of a more general model\cite{mackay}.  Broadly speaking, linear localization corresponds to pure Anderson modes while nonlinear localization to Discrete Breather (DB) modes.  The NNLD provides possibly the simplest system where we may study analytically the competition and coexistence of Anderson modes and discrete breathers.  Using this nomenclature we may designate the region in the ``tie" diagram between the two analytical lines as DB or nonlinear localization region.  The Anderson, or linear localization regime is the region parallel to the $\delta$-axis for small $\zeta$ (except the line at $\delta = 0$). For large nonlinearity and non-degeneracy the linear and nonlinear feature of the localization mixes completely and may not be separated.

\begin{figure}[bt]
\label{NNLD-TD}
\centerline{\psfig{file=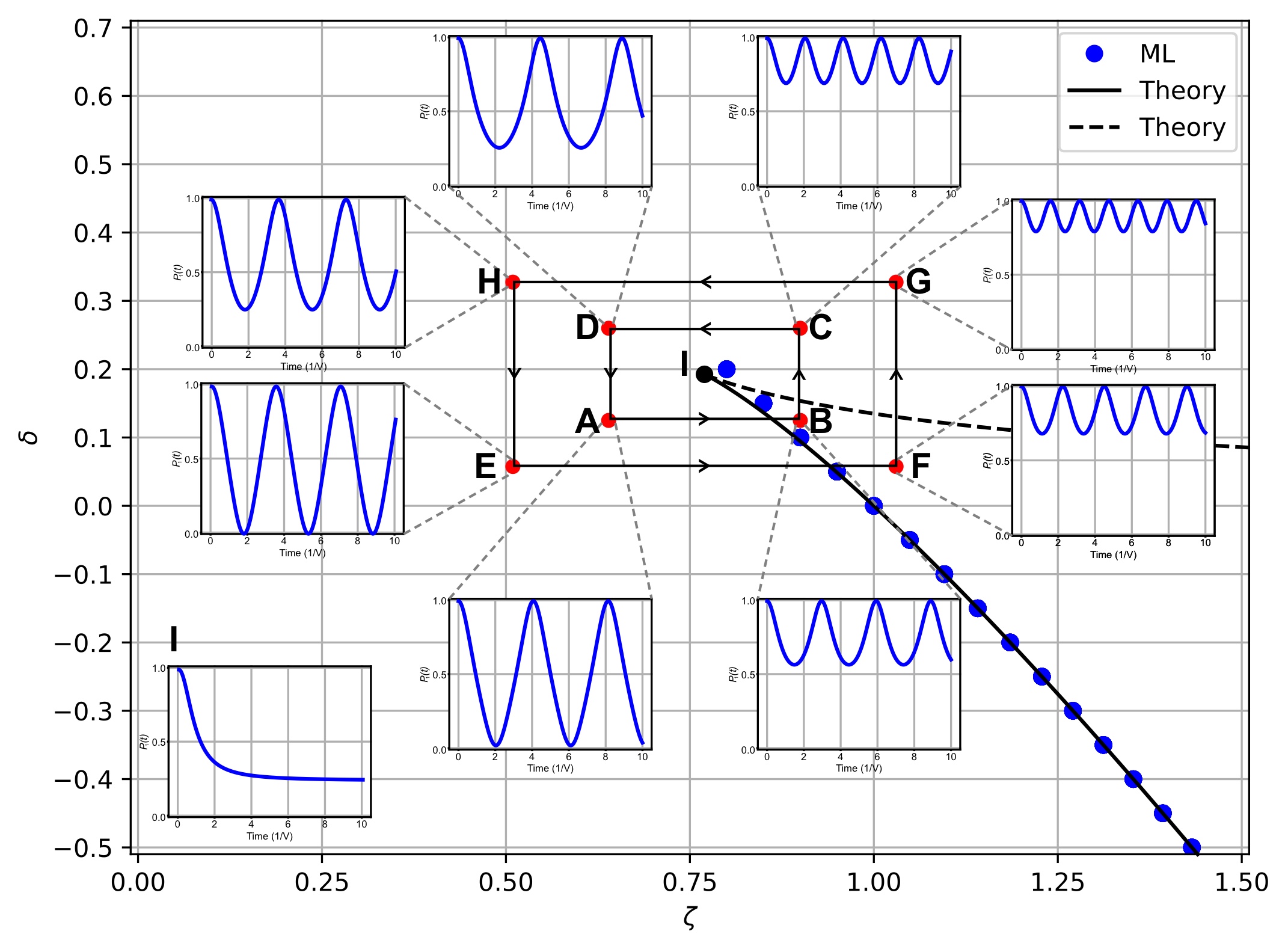,width=5.65in}}
\vspace*{8pt}
\caption{Phase diagram for the self-trapping transition with ML of the  non-degenerate nonlinear dimer with localized initial condition. Time dependent evolution for different parts of the parameter space. The transition across the critical line is discontinuous (A to B) or (E to F) induced by increase in oscillation period. The line terminates in the critical point I where the time dependence is algebraic. We may use paths in the parameter space around the critical point I to go from the free to the self-trapped regime without encountering discontinuity in the character of the evolution.}
\end{figure}

We note in the phase diagram that we may start with an Anderson mode, deform it and reach a DB mode.  This can be done in multiple ways, depending on how we change the parameters $\delta$ and $\zeta$.  One class of paths crosses the self-trapping line discovered by ML while an alternative class may simply reach the same state without crossing it.  The transition from Anderson to DB modes in the present model is similar to a first order phase transition with a critical point.  It is interesting that ML can actually capture the coexistence line that separates the two dynamical regimes in the NNLD.  In Fig. (\ref{NNLD-TD}) we show the actual time dependence of the probability to be in the initially populated site.  We have two sets of paths on the diagram, ABCD and EFGH as well as the central critical point I.  All time dependent curves show the time evolution of the initially populated site.  At A we have almost complete oscillation between the two sites while crossing the critical line we arrive at point B with incomplete, self-trapped motion. The path to C does not change dramatically the localized nature of the motion while the reduction of localization in D is done in a gradual way.  It is clear that crossing the transition line (A to B) results in a discontinuous symmetry breaking although we may reach B also through the continuous line A to D to C to B.  A similar behavior is seen in the second trajectory EFGHE where the passage from E to F is discontinuous while the rest of the trajectory is continuous.

From this analysis we observe additionally  that the crossing of the transition line parallel to the $\zeta$-axis, as for instance from E to F, is done through the increase in period of oscillation that eventually leads to self-trapping.  This is very similar to the degenerate dimer analysis related slowing down induced by nonlinearity\cite{kenkre}.  The transition to an Andreson mode, on the other hand, is a gradual transition to a more and more localized state effected from the increase of the  energy mismatch.  The critical NNLD line terminates in the critical point I, where the time dependence is algebraic\cite{gpt2}.

\section{Conclusions}
Is there any reason to mix physics with AI?  In theoretical physics we have a clear methodology that works quite well. We formulate models, solve them analytically if we can, otherwise we use numerics. What can AI add to this well established  discovering processes? Machine learning can manipulate data and make projections.  If training is done properly we may find hidden properties in data.  In this work we used a well studied model where most results are known analytically.  This aspect provides a great advantage since we may test for any property we wish and be able to compare the ML predictions with solid, analytical results. Furthermore, the specific model of the DNLS equation nonlinear dimer provides a framework for numerous other investigations where the results are only known approximately.

The basic feature of the nonlinear dimer is the self-trapping transition, i.e. a dynamical symmetry breaking that occurs when the nonlinearity passes a certain critical value. This transition is in some sense anticipated in the study of the stationary states of the model where a new state appears along with the analytic continuation of the normal modes of the linear dimer\cite{eilbeck}. Since the equations, however, are nonlinear, knowledge of the nonlinear stationary states cannot give the arbitrary time evolution of the complete problem.  Using an optimization back propagation procedure we saw that we can recover the original self-trapping transition for the extremely localized initial condition\cite{kenkre}.  This is a significant finding because it shows that ML methods are able to distinguish accurately dynamical regimes with different properties.  In addition to the localized initial condition, more general initial conditions give different dynamic dimer evolution; this was also captured by our ML-motivated method.  We note that although ML methods have been applied in equilibrium phase transitions \cite{carleo} the one we studied is {\it dynamic} and thus cannot be accessed through statistical means and forms of image recognition. 

The non-degenerate nonlinear dimer (NNLD) is a simple yet significant system because it includes simultaneously  in the simplest, almost rudimentary form, Anderson localization and discrete breathers at the same time.  The former is due to ``disorder", i.e. the energy mismatch between the two sites while the later is due to self-trapping and the formation of a nonlinear localized mode or DB.  The competition and/or coexistence of the two is an interesting topic of research in general lattice models\cite{mackay}.  The NNLD provides the simplest model that can give clues in this question.  The phase diagram obtained using the precise mathematical criterion of the appearance of a flat region in the effective dimer potential determines essentially three domains\cite{gpt2}. In more modern parlance we could call then Anderson-dominated, DB-dominated and mixed phases. In the first the linear aspect of localization, i.e. the non-degeneracy, dominates. The second is nonlinearity-dominated while in the mixed case the two features are not separated.  It is interesting that similar conclusions were obtained for long nonlinear lattices using more sophisticated mathematical methods\cite{mackay}.  In Table 1 we summarize these aspects for the dimer system.

The use of ML motivated methods was able to determine the NNLD dynamical regimes without resorting to the  precise mathematical condition introduced in the original work.  It is remarkable that this alternative-data driven- approach recovers successfully the analytical results!  This has important consequences since it shows that the range of possibilities for the ML-based discoveries in cases where analytics is not possible is unlimited.   In particular, the knowledge that nonlinear localization proceeds through frequency increase and it is abrupt while the linear one is gradual is a feature that may be directly applied with ML in a possible detection of localized modes.

\begin{table}[ph]   %Table~1 
\tbl{ 
Dynamical behavior of linear and nonlinear dimers discussed in this work}
{\begin{tabular}{@{}llll@{}} \Hline 
\\[-1.8ex] 
TYPE & LINEAR DIMER & NONLINEAR DIMER & TET
 \\[0.8ex] 
\hline \\[-1.8ex] 
Degenerate & Resonant transfer &
Resonant transfer and self-trapping & {} \\ Non-degenerate & Non-resonant transfer &
Linear and nonlinear localization & Effectivelly linear resonant transfer \\[0.8ex] 
\Hline \\[-1.8ex] 
\multicolumn{3}{@{}l}{}\\
\end{tabular}}
\label{tab1}
\end{table}

In the NNLD  there is a mixture of ``static" as well as ``dynamic" energy mismatch. or disorder.  The former stems from the energy difference in the two sites that introduces a natural off-resonance mechanism. The latter comes from the nonlinearity that due to its localization tendency introduces an initial condition dependent non-resonance mechanism.  We noticed that both localization mechanisms work in the same direction and in a sense augment localization be it linear or nonlinear.  There is however a case where the two work in opposition; this is provided by TET.   In the TET configuration the nonlinear terms oppose the Anderson localization feature and as a result we recover the perfect resonance.  This feature is captured fully by an ML method\cite{barmparis}.

\section*{Acknowledgements}

We acknowledge the cofinancing of this research by the European Union and Greek national funds through the Operational Program Crete 2020-2024, under the call ``Partnerships of Companies with Institutions for Research and Transfer of Knowledge in the Thematic Priorities of RIS3Crete," with project title ``Analyzing urban dynamics through monitoring the city magnetic environment" (project KPHP1 - 0029067).

\end{document}